\documentclass[12pt]{article}

\usepackage{titling}
\usepackage[letterpaper, portrait, margin=1in]{geometry}
\usepackage{graphicx}
\usepackage{amsmath}
\usepackage{amssymb}
\usepackage{mathrsfs}
\usepackage{braket}
\usepackage{bm}
\usepackage[T1]{fontenc}
\usepackage{calligra}
\usepackage{array}
\usepackage{caption}
\usepackage{subcaption}
\usepackage{float}
\usepackage{bm}
\usepackage[toc,page]{appendix}
\usepackage[affil-it]{authblk}
\usepackage{mcite}
\usepackage{xcolor}
\usepackage[normalem]{ulem}

\usepackage[colorlinks=true, linkcolor=blue, urlcolor=blue,citecolor=blue]{hyperref}

\hypersetup{colorlinks,linkcolor=blue,citecolor=blue,urlcolor=blue}

\newcommand{\weight}{\mathcal{W}}

\begin{document}
	
	This version of the article has been accepted for publication, after peer review (when applicable) but is not the Version of Record and does not reflect post-acceptance improvements, or any corrections. The Version of Record is available online at: \href{http://dx.doi.org/10.1038/s42005-022-01066-z}{http://dx.doi.org/10.1038/s42005-022-01066-z}
	
	\title{Weak Localisation Driven by Pseudospin-Spin Entanglement}
	
	\author{Frederico Sousa, David T.S. Perkins\textsuperscript{\dag}, Aires Ferreira\textsuperscript{\ddag}}
	\date{}
	\affil{School of Physics, Engineering and Technology and York Centre for Quantum Technologies, University of York, Heslington, York YO10 5DD, United Kingdom}
	\affil{\textnormal{\textsuperscript{\dag}david.t.s.perkins@york.ac.uk, \textsuperscript{\ddag}aires.ferreira@york.ac.uk}}
	\begin{titlingpage}
		{\let\newpage\relax\maketitle}
		
		\begin{abstract}
			At low temperatures, quantum corrections, originating from the interference of the many paths an electron may take between two points, tend to dominate the transport properties of two-dimensional conductors. These quantum corrections increase the resistivity in systems such as two-dimensional electron gases (2DEGs) without spin-orbit coupling (SOC), a phenomenon called weak localisation. Including symmetry-breaking SOC leads to a change from weak localisation (WL) to weak anti-localisation (WAL) of the electronic states, i.e. a WL-to-WAL transition. Here, we revisit the Cooperon, the propagator encoding quantum corrections, within the context of ultra-clean graphene-based van der Waals heterostructures with strong symmetry-breaking Bychkov-Rashba SOC to yield two completely counter-intuitive results. Firstly, we find that quantum corrections vary non-monotonically with the SOC strength, a clear indication of non-perturbative physics. Secondly, we observe the exact opposite of that seen in 2DEGs with strong SOC: a WAL-to-WL transition. This dramatic reversal is driven by mode entanglement of the pseudospin and spin degrees of freedom describing graphene's electronic states. We obtain these results by constructing a non-perturbative treatment of the Cooperon, and observe distinct features in the SOC dependence of the quantum corrections to the electrical conductivity that would otherwise be missed by standard perturbative approaches.
			
		\end{abstract}
	
	\end{titlingpage}
	
	\section*{Introduction}\label{Intro_sec}
	\begin{figure}[t]
		\centering
		\includegraphics[width=\linewidth]{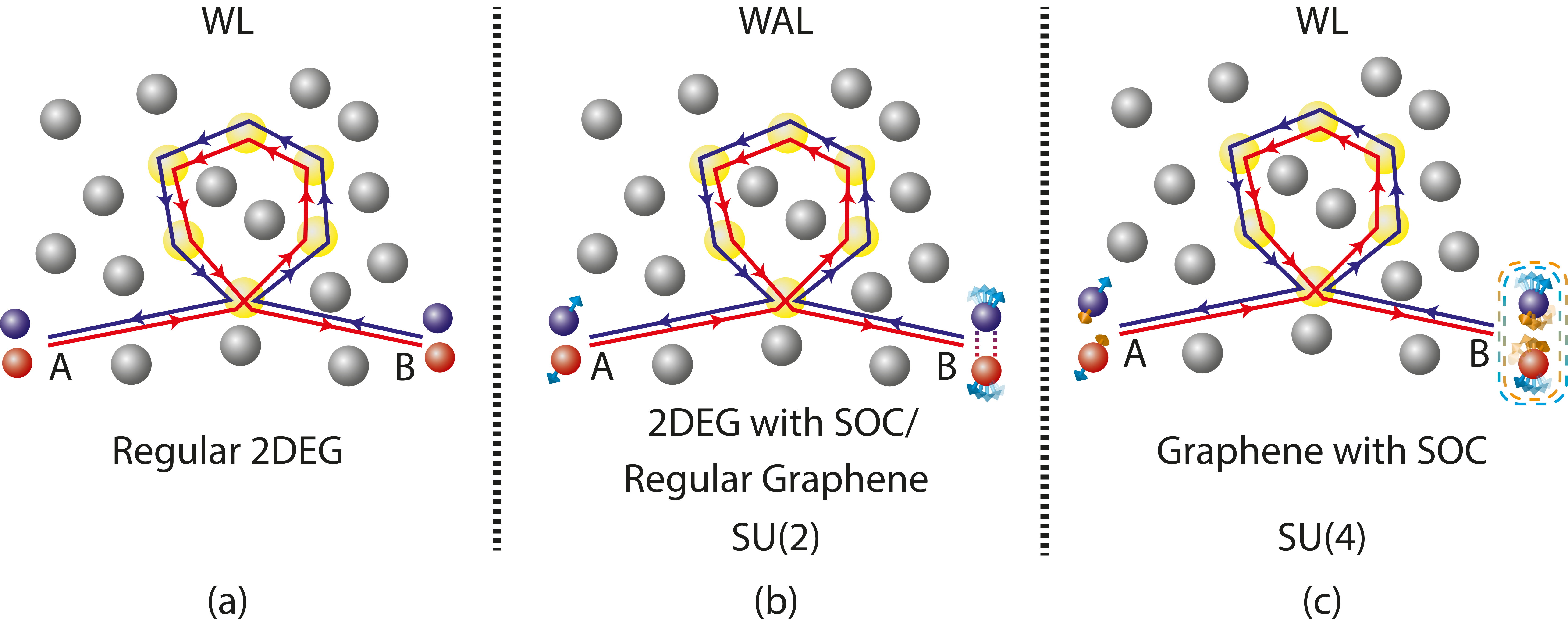}
		\caption{\footnotesize \textbf{Physical interpretation of quantum corrections to the electrical conductivity.} Illustration of a self-intersecting path taken by an electron travelling from point A to point B. The red line indicates the path taken by the electron, whilst the blue line indicates the time-reversed path taken by the electron traversing the loop in the opposite direction to the red path. These are the paths that interfere and lead to quantum corrections. The red and blue spheres represent an electron and hole respectively, whilst the grey spheres denote background impurities in the material. The yellow spheres are the impurities involved in the scattering of the electron. (a): standard 2D electron gas (2DEG), where spin is irrelevant and we observe weak localisation (WL). (b): 2DEG with strong spin-orbit coupling (SOC) or ultra-clean graphene without SOC. The spin/pseudospin is represented by the arrow, and is scrambled upon traversing the loop. We observe weak anti-localisation (WAL) in this situation and find that transport is governed by entangled states of spin/pseudospin. The SU(2) generators are used to model these systems. (c): ultra-clean Rashba-coupled graphene, where the electrons now possesses spin and pseudospin. Both spin-like quantities are scrambled by travelling along the loop, leading to a WL phase. The resulting transport modes are now entangled states of spin and pseudospin. The SU(4) generators govern the physics here.}
		\label{WL_WAL_schematic}
	\end{figure}
	
	Van der Waals (vdW) heterostructures have been the subject of many transport studies over the past two decades \cite{Britnell2012,Georgiou2013,Bertolazzi2013,Geim2013,Gmitra2015,Gmitra2016}, with graphene-based systems being at the centre of these investigations. The excitement surrounding these materials can be attributed to the plethora of novel properties possessed by honeycomb layered materials. For example, the connection between graphene's $\pi$ Berry phase and the absence of coherent backscattering from smooth disorder potentials \cite{Kechedzhi2007} highlights the key role played by the SU(2)-sublattice (pseudospin) degree of freedom. In addition to this, graphene has an exceptionally high spin relaxation length measured at room temperature \cite{Ingla-Aynes2015,Kamalakar_2015,Drogeler2016,Gebeyehu_2019}, making it an ideal gate-tunable high-performance spin-channel for both fundamental science and technology \cite{Avsar2020}.
	
	Recent efforts have focused heavily on spin-orbit effects in transport and magnetism, including all-electrical spin-to-charge/charge-to-spin conversion processes  \cite{Safeer2019,Ghiasi2019,Benitez2020,Cavill2020,Li2020}, and current-induced spin-orbit torque \cite{Sousa_2020,Zollner_2020,Hidding_2021,deSousa_2021}, in light of the technological potential of spintronics \cite{Manchon_2019}. These effects are particularly prominent in ultra-thin two-dimensional (2D) Dirac materials with proximity induced Bychkov-Rashba spin-orbit coupling (SOC) due to the interfacial breaking of inversion symmetry \cite{Rashba1984}. One of the most promising materials with technological applications are those comprised of graphene and transition metal dichalcogenides (TMDs), due to their ability to enhance the SOC strength without jeopardising the Dirac nature of graphene's electronic states \cite{Gmitra2016,Wang2015,Avsar2014}.
	
	Rashba-coupled graphene heterostructures exhibit interesting properties in both their electrical and spin transport. Most notably for this paper, typical disordered Rashba-coupled graphene has been shown to house a weak anti-localisation (WAL) phase due to the strong SOC induced by its TMD partner \cite{Wang2016,Yang2016,Yang2017,Volkl2017,Wakamura2018}. This is in contrast to isolated disordered graphene monolayers, whose intrinsic SOC is negligible (typically $\sim 40\, \mu$eV \cite{Sichau2019}), which instead exhibit a weak localisation phase \cite{Morozov2006,Tikhonenko2008,Ki2008,Pezzini2012}. This WL-to-WAL transition in the quantum corrections, $\Delta \sigma$, to the electrical conductivity, $\sigma$, has been discussed by Imura et. al. \cite{Imura2010}, and therefore lines up with the picture painted by the numerous studies of 2D electron gases (2DEGs) throughout the late 20\textsuperscript{th} century, see the work of Bergmann \cite{Bergmann1984} and the references therein.
	
	In a regular 2DEG with no SOC, electrons exhibit WL due to the constructive interference of time reversed self-intersecting paths, an example is shown in Fig. \ref{WL_WAL_schematic}a. These paths lead to an increased probability of an electron appearing at the point of self-intersection, meaning that the electron becomes weakly localised around this point. Consequently, the probability of the electron being measured at point B is reduced and thus the electrical conductivity, a measure of the number of electrons travelling across the system from A to B, is reduced, $\Delta \sigma <0$. Observations of WL have been made in several experiments \cite{Dolan1979,Hoffmann1982,Abraham1983,Pratumpong2000}, and explained in multiple theoretical works \cite{Abrahams1979,Gorkov1979,Abrahams1980}.
	
	If we now include strong SOC into the 2DEG, we find that the system exhibits WAL instead. This can be attributed to the correlation between momentum and spin enforced by Rashba SOC \cite{Rashba_2009}, which leads to the statistical scrambling of the electron's spin when traversing the loop of these paths (see Fig. \ref{WL_WAL_schematic}b), and thus results in destructive interference. Clearly, the electron now has a reduced probability of being measured at the self-intersection point, and hence an increased chance of being observed at B. Therefore, the quantum corrections are positive, $\Delta \sigma > 0$. We may think of this as scrambling spin produces a minus sign in the quantum correction. Here the resulting transport can be tied to effective spin singlets and triplets, i.e. states of entangled electrons. This therefore implies a WL-to-WAL transition as the SOC strength is increased. The WAL phase has been observed in a multitude of materials \cite{Sharvin1981,Bergmann1982a,Knap1996,Koga2002}. Theoretical interpretations of this switching from WL to WAL can be found in the seminal works of Bergmann \cite{Bergmann1982b,Bergmann1983} and Hikami et. al. \cite{Hikami1980}, whilst direct observation has been made by Caviglia et. al. \cite{Caviglia2010}.
	
	For the case of ultra-clean graphene (i.e. with reduced intervalley scattering due to the absence of point-like defects \cite{Engels2014,Couto2014,Joucken2021}) without SOC, we observe the same WAL behaviour as the SOC 2DEG in both theory and experiment \cite{Suzuura2002,McCann2006,Wu2007,Tikhonenko2009}. The intrinsic SOC in graphene is negligible and so cannot be responsible for this observation of WAL. Rather, graphene generates another spin-like quantum number for the electrons called pseudospin, due to being formed of two interwoven sublattices. It turns out that the pseudospin is the quantity scrambled by travelling around the loops which leads to WAL. In analogy to the SOC 2DEG, the states of interest are effective pseudospin singlets and triplets.
	
	In this paper, we focus on the inclusion of strong Rashba SOC into ultra-clean graphene. We prove that the quantum correction generated is localising in nature, and observe a WAL-to-WL transition upon increasing the SOC strength in our theoretical analysis. This is in contrast to the WL-to-WAL transition driven by increasing SOC strength in conventional metals \cite{Dolan1979,Hoffmann1982,Abraham1983,Pratumpong2000,Abrahams1979,Gorkov1979,Abrahams1980,Sharvin1981,Bergmann1982a,Knap1996,Koga2002,Bergmann1982b,Bergmann1983,Hikami1980,Caviglia2010}. Intuitively, we may understand the WAL-to-WL transition predicted in this work in terms of the pseudospin and spin degrees of freedom. When traversing a loop, both spin-like quantities are scrambled, see Fig. \ref{WL_WAL_schematic}c, and so each produce a minus sign. The product of these minus signs cancel and so the system displays a WL phase.
	
	To accurately include SOC into our analysis of the electrical conductivity, we find that large momentum contributions to the diffusive particle-particle propagator (the Cooperon), which governs the quantum corrections, become relevant for strong SOC. Therefore, unlike in previous works \cite{Ochoa2014,Ilic2019}, where perturbative methods were used to focus purely on small momentum contributions to the Cooperon, we generalise the non-pertubative approach of Wenk et. al. \cite{Wenk2010} used for 2DEG systems for use in graphene to evaluate the Cooperon exactly.
	
	\section*{Results}\label{Results_sec}
	\subsection*{Disorder and Quantum Interference Corrections} \label{Cooperon_theory_subsec}
\begin{figure}[t]
		\centering
		\includegraphics[width=0.8\linewidth]{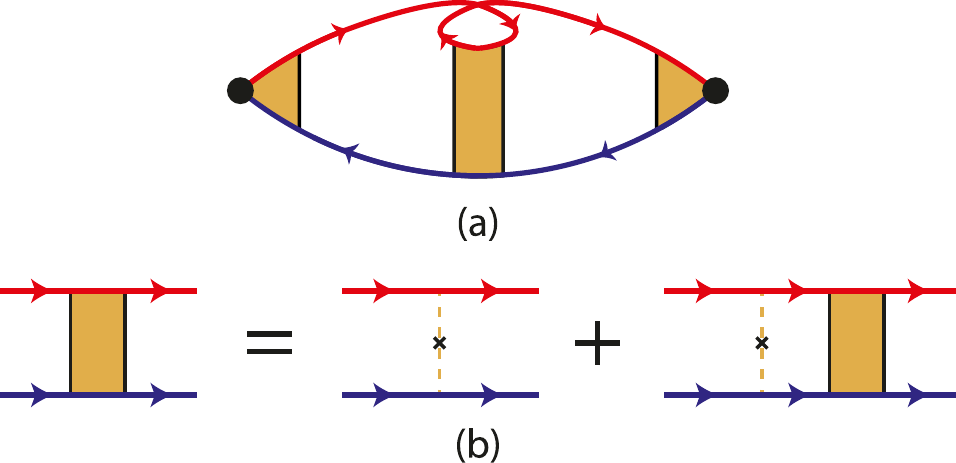}
		\caption{\footnotesize \textbf{Feynmann diagrams describing quantum corrections.} (a): Diagrammatic representation of the quantum correction's linear response function, including vertex corrections. The red/blue lines denote retarded/advanced disorder-averaged Green's functions, the black circles represent the current vertices, and the central orange shaded region represents the Cooperon. (b): Diagrammatic series for the Cooperon. The orange dashed lines represent scattering from the impurity located at the cross.}
		\label{WL_Cooperon_diagram}
	\end{figure}

	Using standard diagrammatic techniques \cite{AGD,Mahan} and methods of disorder averaging \cite{Bruus_Flensberg}, we determine the corrections to the DC electrical conductivity due to quantum interference by calculating the Langer-Neal diagram \cite{Langer1966}, Fig. \ref{WL_Cooperon_diagram}a, within the semi-classical regime, $\varepsilon \tau_{p} \gg 1$ ($\varepsilon$: Fermi energy, $\tau_{p}$: momentum relaxation time). This diagram yields the leading order correction \cite{Araki2014}
	\begin{equation}
		\Delta\sigma = \frac{e^{2}}{2\pi \mathcal{V}_{d}} \, \text{tr}\left[ \sum_{\mathbf{k},\mathbf{q}} \weight^{\beta\alpha}_{\gamma\delta}(\mathbf{k},-\mathbf{k}) C^{\alpha\beta}_{\delta\gamma}(\mathbf{q}) \right],
		\label{WL_conductivity_general_relation}
	\end{equation}
	where
	\begin{equation}
	    \mathcal{W}^{\beta\alpha}_{\gamma\delta}(\mathbf{k},\mathbf{k}') = [\mathcal{G}^{A}(\mathbf{k}) \tilde{v}_{x} \mathcal{G}^{R}(\mathbf{k})]_{\gamma\alpha} [\mathcal{G}^{R}(\mathbf{k}') \tilde{v}_{x} \mathcal{G}^{A}(\mathbf{k}')]_{\beta\delta},
	\end{equation}
	is the weight matrix, $\mathcal{G}^{R/A}(\mathbf{k})$ are the retarded/advanced disorder-averaged Green's functions at the Fermi surface, $\tilde{v}_{x}$ is the renormalised velocity operator along the $x$-axis, and $\mathcal{V}_{d}$ is the system's $d$-dimensional volume. The Cooperon, $C(\mathbf{q})$, represented diagrammatically in Fig. \ref{WL_Cooperon_diagram}b, can be written in a more compact form by using a matrix basis such that
	\begin{equation}
		C(\mathbf{q}) = \frac{\Gamma_{0}}{1 - \Gamma_{0} \Pi(\mathbf{q})},
		\label{Cooperon}
	\end{equation}
	where $\Gamma_{0} = n U \otimes U$, $U$ describes the impurity landscape (we shall discuss this shortly), $n$ is the impurity number density, and $\Pi(\mathbf{q})$ is the Cooperon self-energy. The transport modes are then just eigenstates of the \textit{Cooperon Hamiltonian}, $H_{C} = 1 - \Gamma_{0} \Pi(\mathbf{q})$, with eigenvalues $E_{c}(\mathbf{q})$ (see refs. \cite{Wenk2010,Ferreira2021} for further discussion). The dominance of a given mode may be understood in terms of any gaps appearing in its dispersion relation (eigenvalue); gapless modes, and modes with suitably small gaps, will lead to the largest conductivity corrections. We refer to this set of states as the \textit{transport basis}.
	
	To understand whether a transport mode is localising or delocalising in nature however, we need to analyse the momentum integrated weight matrix. We obtain this information by inspecting the sign of the diagonal elements of this matrix when written in the transport basis: a positive element means WAL, whilst a negative element means WL. This is extremely useful when trying to understand which states dominate the transport phenomena and why transitions between WL and WAL occur.
	
	
	As mentioned above, the details of the impurity landscape are contained in the matrix structure of $U$: each type of scattering has a specific matrix form associated to it, along with a characteristic strength. For example, we can include the effects of intervalley scattering induced by a hollow position adatom via a term that is off-diagonal in the valley basis, $-u_{x} \tau_{x} \otimes \sigma_{z} \otimes s_{0}$ \cite{Pachoud2014}. Further complexity can be introduced with the inclusion of a sublattice disorder term, which corresponds to offsetting the A and B sublattice energies from one another by $2u_{z}$, $u_{z} \tau_{z} \otimes \sigma_{z} \otimes s_{0}$ into $U$ (the matrices $\tau_{i},\sigma_{i},$ and $s_{i}$ are defined in the following section). Clearly, many types of impurity scattering mechanisms may be included in models of disordered graphene, we therefore refer the reader to refs. \cite{Pachoud2014,Basko2008,McCann2012} for further examples.
	
	In many previous studies \cite{Hikami1980,Suzuura2002,McCann2006,Ochoa2014,Ilic2019,Araki2014}, the Cooperon self-energy has been calculated perturbatively, $\Pi(\mathbf{q}) \simeq \Pi(\mathbf{0}) + q \Pi'(\mathbf{0}) + q^{2}\Pi''(\mathbf{0})/2$. However, in the case of strong SOC effects, $\Pi(\mathbf{q})$ cannot be handled using such methods, as minima appear in the dispersion relations of the transport modes away from $q = 0$, leading to gaps small enough to generate noteworthy conductivity corrections. Consequently, we must make use of the non-perturbative formula (see \href{http://dx.doi.org/10.1038/s42005-022-01066-z}{Supplementary Note 1})
	\begin{equation}
		\Pi(\mathbf{q}) = i \int_{0}^{2\pi} \frac{d\theta}{8\pi} \widetilde{\rho}(\varepsilon) \left[ \frac{1}{\mathcal{H}^{(2)}_{\mathbf{k}_{F}} - \mathcal{H}^{(1)}_{\mathbf{q}-\mathbf{k}_{F}} + 2i\eta} - \frac{1}{\mathcal{H}^{(1)}_{\mathbf{k}_{F}} - \mathcal{H}^{(2)}_{\mathbf{q}-\mathbf{k}_{F}} - 2i\eta} \right],
		\label{Cooperon_self_energy_non_perturbative}
	\end{equation}
	where $\widetilde{\rho}(\varepsilon)$ is the average DOS of the Rashba-split bands, $\widetilde{\rho} = (\rho_{+}+\rho_{-})/2$, $\rho_{\pm}$ is the DOS of the spin minority/majority band, $\mathcal{H}^{(1)}_{\mathbf{k}} = H_{0\mathbf{k}} \otimes \mathcal{I}_{D}$, $\mathcal{H}^{(2)}_{\mathbf{k}} = \mathcal{I}_{D} \otimes H_{0\mathbf{k}}$, $\mathcal{I}_{D}$ is the $D \times D$ identity matrix, $D$ is the matrix dimension of the Hamiltonian, and $\eta$ is the quasiparticle broadening related to the scattering rates of different processes within the disordered material. Eq. (\ref{Cooperon_self_energy_non_perturbative}) is a generalisation of the expressions implemented for 2DEGs \cite{Wenk2010}, allowing for the capture of the SU(2) nature of graphene's sublattice in an entirely non-perturbative fashion.
	
	Now, as a matter of formality, we must address the behaviour of the momentum integral of the Cooperon (i.e. the integral over $\mathbf{q}$ in eq. \ref{WL_conductivity_general_relation}), which suffers from both ultra-violet and infra-red divergences. To handle this, we introduce natural cutoffs set by the mean free path, $l$, and the inelastic scattering length (sometimes referred to as the phase coherence length), $l_{\phi} \gg l$. The upper and lower limits of the momentum integral are then simply $l^{-1}$ and $l_{\phi}^{-1}$ respectively. However, by working at sufficiently low temperatures, $l_{\phi}$ will exceed the system size, $L$. This is the case we work within and so our infra-red cutoff is set as $L^{-1}$.
	
	\subsection*{Rashba-Coupled Graphene} \label{Sec_Rashba}
	We now apply the above method for calculating the conductivity corrections to Rashba-coupled graphene. In the absence of disorder, this system has the effective Hamiltonian
	\begin{equation}
		H_{0\mathbf{k}} = v \tau_{0} \otimes \boldsymbol{\sigma} \cdot \mathbf{k} \otimes s_{0} + \alpha \tau_{0} \otimes (\sigma_{1} \otimes s_{2} - \sigma_{2} \otimes s_{1}),
	\end{equation}
	where $v$ is the Fermi velocity of the massless Dirac fermions, $\alpha$ is the Rashba SOC strength, and $\tau_{i}$, $\sigma_{i}$, and $s_{i}$ ($i = 0,1,2,3$) are the Pauli matrices acting on the valley (isospin), sublattice (pseudospin), and spin degrees of freedom (DOFs), respectively. This Hamiltonian is written in the \textit{valley basis} $\ket{\psi} = (KA\uparrow,KA\downarrow,KB\uparrow,KB\downarrow,-K'B\uparrow,-K'B\downarrow,K'A\uparrow,K'A\downarrow)^{T}$, with $K$ and $K'$ as the valley indices, and $A$ and $B$ as the sublattice indices.
	
	We include disorder in our model via the real-space impurity term,
	\begin{subequations}
	\begin{equation}
		H_{\text{imp}} = \sum_{i} U \delta^{(2)}(\mathbf{r}-\mathbf{r}_{i})
	\end{equation}
	\begin{equation}
	    U = u_{0} \tau_{0} \otimes \sigma_{0} \otimes s_{0} - u_{x}\tau_{x} \otimes \sigma_{z} \otimes s_{0} - u_{y} \tau_{y} \otimes \sigma_{z} \otimes s_{0},
	\end{equation}
	\end{subequations}
	where the sum is over all impurities, with $\mathbf{r}_{i}$ being the position of the $i$\textsuperscript{th} impurity. The complete real-space Hamiltonian is thus $H = (H_{0\mathbf{k}})_{\mathbf{k}\rightarrow \mathbf{p}} + H_{\text{imp}}$, with $\mathbf{p}$ the 2D momentum operator. The first term of $U$ is responsible for scattering within a valley, yielding the intravalley scattering rate $\tau_{v}^{-1} = nu_{0}^{2}\varepsilon/(4v^{2})$, where $\varepsilon$ is the Fermi energy. In contrast, the second and third terms allow for scattering between valleys, resulting in the intervalley scattering rate $\tau_{iv}^{-1} = n(u_{x}^{2}+u_{y}^{2})\varepsilon/(4v^{2})$. The sum of these scattering rates gives the quasiparticle broadening, $\eta = \tau_{v}^{-1} + \tau_{iv}^{-1}$.
	
	We shall initially focus our attention on the case of ultra-clean graphene where intervalley scattering is negligible, as this yields the WAL-to-WL transition (the role of intervalley processes will be addressed later on). Consequently, $C(\mathbf{q})$ and $\mathcal{W}(\mathbf{k},-\mathbf{k})$ are 16-dimensional square matrices, which can be written in terms of 15 SU(4) generators  $\gamma_{a0} = \sigma_{a} \otimes s_{0}$ ($a=x,y,z$), $\gamma_{0a} = \sigma_{0} \otimes s_{a}$ ($a=x,y,z$) and $\gamma_{ab} = \sigma_{a} \otimes s_{b}$ ($a,b=x,y,z$), supplemented with the $4\times4$ identity matrix $\gamma_{00} = \sigma_{0} \otimes s_{0}$. In general, these generators can be used to describe many useful transport properties within graphene besides just the electrical conductivity, such as the spin-galvanic susceptibility and spin conductivity tensor \cite{Ferreira2021}. Any spin-charge perturbation may be incorporated into the Hamiltonian by coupling to the appropriate $\gamma_{ij}$.
	
	Finally, to understand how entanglement manifests in graphene with strong Rashba SOC, it will be more convenient to work in terms of an alternate basis, the $\textit{decoupled basis}$, to that of $\ket{\psi} \otimes \ket{\psi}$. We define this new basis in terms of the effective singlet, $\ket{s}$, and triplets, $\ket{t_{i}}$, of the spin and pseudospin, similar to that of Araki et. al. \cite{Araki2014}. Using spin to illustrate these states, we write
	\begin{equation}
		\begin{pmatrix}
			 \ket{t_{1}} \\ \ket{s} \\ \ket{t_{2}} \\ \ket{t_{3}}
		\end{pmatrix}
		=
		\frac{1}{\sqrt{2}}
		\begin{pmatrix}
			1 & 0 & 0 & 1 \\
			0 & 1 & -1 & 0 \\
			0 & 1 & 1 & 0 \\
			1 & 0 & 0 & -1
		\end{pmatrix}
		\begin{pmatrix}
			\ket{\uparrow\uparrow} \\ \ket{\uparrow\downarrow} \\ \ket{\downarrow\uparrow} \\ \ket{\downarrow\downarrow}
		\end{pmatrix}.
	\end{equation}
	The singlets and triplets for pseudospin can be obtained from this by making the substitutions $\uparrow \rightarrow A$ and $\downarrow \rightarrow B$. The decoupled basis is thus defined as
	\begin{equation}
	    \ket{\Psi} = (ss, \, st_{1}, \, st_{2}, \, st_{3}, \, t_{1}s, \, t_{2}s, \, t_{3}s, \, t_{1}t_{1}, \, t_{1}t_{2}, \, t_{1}t_{3}, \, t_{2}t_{1}, \, t_{2}t_{2}, \, t_{2}t_{3}, \, t_{3}t_{1}, \, t_{3}t_{2}, \, t_{3}t_{3})^{T}
	\end{equation}
	where the $T$ denotes transposition, and we have used $ab = \ket{a} \otimes \ket{b}$ for notational convenience. The first letter appearing in each component refers to the pseudospin state, whilst the second letter refers to the spin state.
	
	\subsection*{Weak Rashba SOC} \label{Sec_weak_SOC}
	To illustrate how spin and pseudospin begin to become entangled, let us first consider the limit of weak SOC, $\alpha \ll \eta$. We choose to focus on the weight matrix for this, due to its simple form compared to the Cooperon. Specifically, we obtain
	\begin{equation}
		W_{\text{MI}} = \sum_{\mathbf{k}} \mathcal{W}(\mathbf{k},-\mathbf{k}) = \begin{pmatrix}
			W_{11} & 0 & W_{13} & 0 \\
			0 & -\frac{W_{11}}{4} & 0 & 0 \\
			-W_{13} & 0 & -\frac{3W_{11}}{4} & W_{34} \\
			0 & 0 & -W_{34} & 0
		\end{pmatrix},
		\label{Weak_SOC_weight_matrix}
	\end{equation}
	in the decoupled basis, where each element, $W_{ij}$, is a $4\times4$ matrix. We find that $W_{11}$ is diagonal and proportional to $\varepsilon\eta^{-3}$, whilst the off-diagonal components $W_{13},W_{34} \propto \alpha\eta^{-3}$. For the interested reader, the exact forms of the $W_{ij}$ are given in the \href{http://dx.doi.org/10.1038/s42005-022-01066-z}{Supplementary Note 2}. Hence, we can justify the common approximation of ignoring the matrix structure of $\mathcal{W}$ and taking a weighted trace of the Cooperon to obtain $\Delta\sigma$ \cite{Ochoa2014,McCann2012,Lu2011}. However, this is not true in the strong SOC case, where the off-diagonal elements of the weight matrix acquire significant magnitudes.
	
	Setting $\alpha = 0$ we note that the off-diagonal components of Eq. (\ref{Weak_SOC_weight_matrix}) vanish, meaning that the diagonal basis of $W_{\text{MI}}$ is simply the decoupled basis in the absence of Rashba SOC. Upon the inclusion of a weak Rashba coupling, we find that the diagonal basis of $W_{\text{MI}}$ cannot be written in terms of simple product states. The same is also true for the Cooperon [Eq. (\ref{Cooperon})], meaning that the transport modes must be formed of states entangling the spin and pseudospin DOFs of the electrons.
	
	The largest quantum corrections are given by the eigenstates of the Cooperon Hamiltonian with a small gap or no gap at all. Hence, the most important states and their respective gaps for weak Rashba coupling are,
	\begin{subequations}
	\begin{align}
		&\ket{\phi_{ss}} = ss \, ; \, \Delta_{ss} = 0,
		\label{Weak_SOC_ss_state} \\
		&\ket{\phi_{st_{1}}} = st_{1} - 2\alpha\tau_{p} t_{2}t_{3} \, ; \, \Delta_{st_{1}} = \frac{\tau_{p}}{2\tau_{\textrm{DP}}},
		\label{Weak_SOC_st1_state} \\
		&\ket{\phi_{st_{2}}} = st_{2} - 2\alpha\tau_{p} t_{2}t_{3} \, ; \, \Delta_{st_{2}} = \frac{\tau_{p}}{2\tau_{\textrm{DP}}},
		\label{Weak_SOC_st2_state} \\
		&\ket{\phi_{st_{3}}} = st_{3} - 2\alpha\tau_{p} (t_{1}t_{2} + t_{2}t_{1}) \, ; \, \Delta_{st_{3}} = \frac{\tau_{p}}{\tau_{\textrm{DP}}},
		\label{Weak_SOC_st3_state}
	\end{align}
	\label{Weak_SOC_states}%
	\end{subequations}
	where $\tau_{p}^{-1} = 2\eta$ is the momentum relaxation rate, and $\tau_{\textrm{DP}}^{-1} = 4\alpha^{2}\tau_{p}$ is the Dyakonov-Perel relaxation rate \cite{Offidani2018}, which clearly satisfies $\tau_{\textrm{DP}}^{-1} \ll \tau_{p}^{-1}$ in the weak SOC limit. Here we have given the zero-momentum form of the transport modes for sake of brevity. All other eigenstates of the Cooperon have notably larger gaps, and so are sub-dominant.
	
	From the states listed in Eq. (\ref{Weak_SOC_states}), we see that the gapless pseudospin singlet state, which dominates the quantum corrections in the absence of SOC, splits into four separate modes. These modes can be identified by their dominant contribution: $ss$ in Eq. (\ref{Weak_SOC_ss_state}), and $st_{i}$ for Eqs. (\ref{Weak_SOC_st1_state})-(\ref{Weak_SOC_st3_state}). The latter three modes exhibit the inescapable entanglement of spin and pseudospin, even at zero momentum, though they can be approximated as $st_{i}$ product states due to the weak mixing of the $t_{j}t_{k}$ products courtesy of their $\alpha\tau_{p}$ prefactor. Hence, transport in weak Rashba-coupled graphene can be described approximately in terms of pseudospin-spin product states.
	
	With this in mind, the sign of the quantum corrections due to each of the states listed above can be understood from the matrix element
	\begin{equation}
		W_{11} = \frac{\varepsilon}{4\eta^{3}} \begin{pmatrix}
			-1 & 0 & 0 & 0 \\
			0 & 1 & 0 & 0 \\
			0 & 0 & 1 & 0 \\
			0 & 0 & 0 & 1
		\end{pmatrix}.
	\end{equation}
	From this we see that the $ss$ state leads to WL, whilst the $st_{i}$ states encourage WAL. Despite the $ss$ mode being gapless, we find that the $st_{i}$ gaps are sufficiently small that their sum dominates over the $ss$ contribution, leading to a WAL phase for weak SOC (i.e. $\Delta\sigma > 0$). This result agrees with previous works where the Cooperon was handled in a perturbative manner \cite{McCann2012,Ilic2019}.
	
	\begin{figure}[t]
		\centering
		\includegraphics[width=0.4\linewidth]{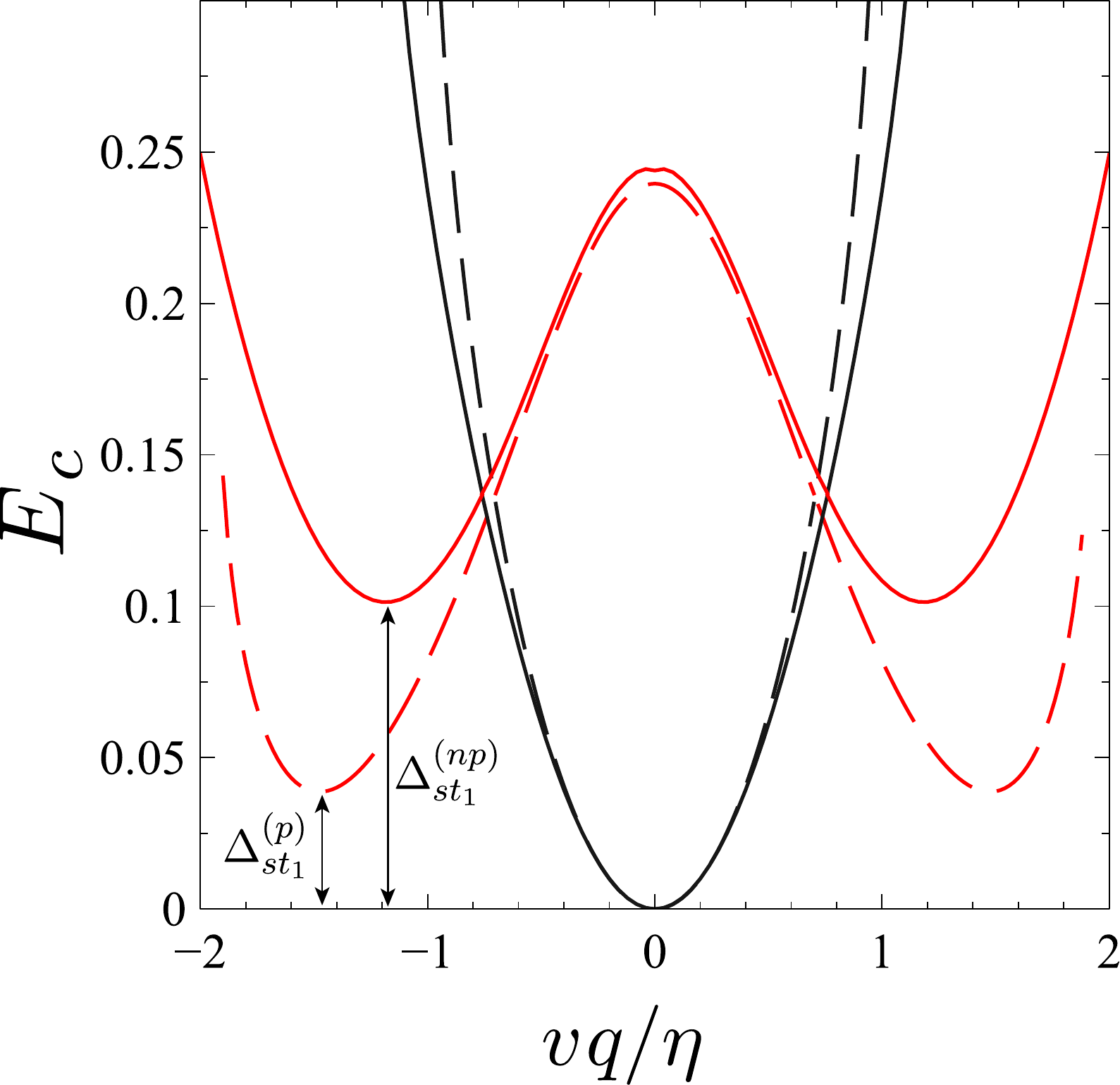}
		\caption{\footnotesize \textbf{Perturbative versus non-perturbative treatment of the Cooperon Hamiltonian's dispersion relations.} Comparison of the $ss$ (black) and $st_{1}$ (red) transport modes' dispersion relations, $E_{c}(q)$, calculated via perturbative (dashed) and non-perturbative (solid) methods. We see that significant minima develop in the $st_{1}$ mode away from the origin, which are underestimated by perturbative approaches ($\Delta_{st_{1}}^{(np)} > \Delta_{st_{1}}^{(p)}$). Here we used $\varepsilon = 0.2$\,eV, $\eta = 0.7\,$meV, and $\alpha = 2\eta/3$ for the Fermi energy, quasiparticle broadening, and Rashba coupling strength respectively.}
		\label{Strong_SOC_dispersion_comparision}
	\end{figure}
	
	\subsection*{Strong Rashba SOC} \label{Sec_strong_SOC}
	Turning our attention towards the lesser understood limit of strong Rashba coupling, $\alpha \gtrsim \eta$, we obtain a far more complex structure for the momentum integrated weight matrix,
	\begin{equation}
		W_{\text{MI}} = \sum_{\mathbf{k}} \mathcal{W}(\mathbf{k},-\mathbf{k}) = \begin{pmatrix}
			W_{11} & W_{12} & W_{13} & W_{14} \\
			W_{12}^{\dagger} & W_{22} & W_{23} & W_{24} \\
			W_{13}^{\dagger} & W_{23} & W_{33} & W_{12}^{\dagger} \\
			W_{14} & W_{24}^{\dagger} & W_{12} & W_{44}
		\end{pmatrix},
    \end{equation}
	with the individual matrix elements given in \href{http://dx.doi.org/10.1038/s42005-022-01066-z}{Supplementary Note 2}. The off-diagonal elements are still proportional to $\alpha\eta^{-3}$, which is no longer a small parameter, and hence cannot be ignored. As in the weak SOC limit, the same is true for the Cooperon, thus the transport modes in Rashba-coupled graphene are inherently entangled states of spin and pseudospin.
	
	We find that the majority of the transport modes become entangled in the strong SOC limit (see \href{http://dx.doi.org/10.1038/s42005-022-01066-z}{Supplementary Note 3}). The dispersion relations of the entangled states acquire significant minima far away from the origin, see Fig. \ref{Strong_SOC_dispersion_comparision}, leading perturbative approaches to underestimate the gap size at non-zero $\mathbf{q}$. We therefore use Eq. (\ref{Cooperon_self_energy_non_perturbative}) to obtain an accurate expression for the Cooperon when $vq \approx \eta$. We note that at zero momentum, the $ss$ state mentioned in the weak SOC limit survives as the only gapless mode, and remains localising in nature. Using this non-pertubative method for calculating the Cooperon, we may determine the quantum corrections to the DC conductivity accurately for strong Rashba coupling.
	
\begin{figure}[t]
		\centering
		\includegraphics[width=0.7\linewidth]{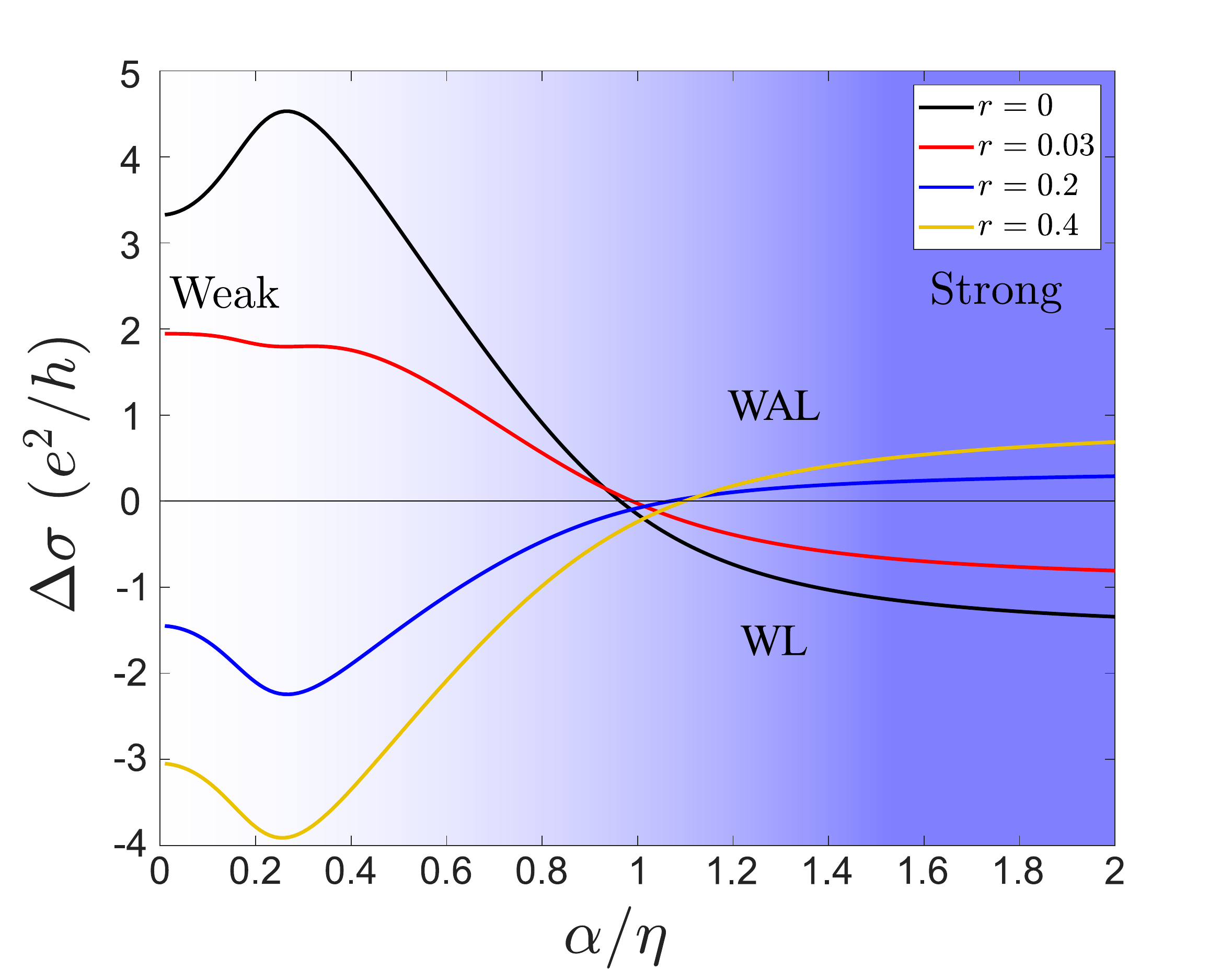}
		\caption{\footnotesize \textbf{Quantum corrections to the electrical conductivity.} Quantum corrections to the electrical conductivity, $\Delta\sigma$, as a function of Rashba coupling strength for various ratios of the intervalley and intravalley scattering rates, $r = \tau_{v}/\tau_{iv}$ ratios (keeping $\eta$ fixed). WAL-to-WL transitions happen for ultra-clean samples ($r \lesssim 0.03$), whilst WL-to-WAL transitions occur in dirty samples ($r \gtrsim 0.2$). The background colour gradient indicates the shift from the weak SOC regime to the strong SOC regime. Here we used $\varepsilon = 0.2$\,eV, $L = 1\,\mu$m, $\eta \simeq 1.2$\,meV (corresponding to a mean free path of $l = v\tau_{p} \simeq 0.3\,\mu$m, with $v = 10^{6}\,\text{ms}^{-1}$), and $n = 10^{12}\,\text{cm}^{-2}$ for the Fermi energy, linear length, quasiparticle broadening, and impurity density respectively, whilst also assuming that $u_{x} = u_{y}$ (with $\eta$ fixed) for the intervalley scattering potentials.}
		\label{Quantum_corrections_plots}
	\end{figure}

	Handling these matrices and their product numerically (see \cite{Zenodo} for the associated code), we plot the quantum corrections to the conductivity as a function of $\alpha$ in Fig. \ref{Quantum_corrections_plots} for various intervalley scattering rates. Here we see that in the case of ultra-clean graphene (black line) the system initially becomes less localised as $\alpha$ is increased, before transitioning from WAL to WL at $\alpha \approx \eta$. We can understand this behaviour by looking at the gaps of the transport modes and appreciating whether they localise or delocalise the electrons. The initial push towards anti-localisation with the increase of $\alpha$ is a result of gaps appearing in the dispersion relations of the entangled states away from zero momentum, see Eq. (\ref{Weak_SOC_states}) and Fig. \ref{Strong_SOC_dispersion_comparision}. Consequently, despite being the only gapless mode, the $ss$ state becomes more sub-dominant to the entangled states for weak to intermediate SOC strengths. However, by further increasing the Rashba coupling strength we find that the gaps of the entangled states begin to grow and become too large to dominate over the gapless $ss$ state, pushing the system towards a more localised state. At the point $\alpha \simeq  \eta$, the gapless $ss$ mode begins to dominate over the entangled states, thus yielding a WL phase for $\alpha \gtrsim  \eta$.
	
	Finally, from the various intervalley scattering rates used in Fig. \ref{Quantum_corrections_plots}, we can see that an increase in the intervalley scattering rate leads to another reversal of the transition, suppressing the WAL-to-WL switching in favour of WL-to-WAL. Clearly, an ultra-clean sample with minimal intervalley scattering will therefore be needed to observe this SOC driven WAL-to-WL transition. The exact value of $\tau_{iv}^{-1}$ at which the transition reverses is not a trivial point to determine in a general manner given the size of expressions generated by this $64 \times 64$ matrix formalism. In addition to the intervalley scattering rate, other system parameters such as the carrier concentration and mean free path also play an important role. Studies of how $\Delta\sigma$ depends upon $\varepsilon$ and $l/L$ can be found in \href{http://dx.doi.org/10.1038/s42005-022-01066-z}{Supplementary Note 4}. We also note that at high enough temperatures, the infrared cut-off appearing in the momentum integration of the Cooperon will change from $L^{-1}$ to $l_{\phi}^{-1}$, where $l_{\phi}^{\null}$ is the phase relaxation length \cite{Abrahams1979,Gorkov1979,Abrahams1980}.

	\section*{Discussion}\label{Discussion_sec}
	The WAL-to-WL transition reported above is in stark contrast to the WL-to-WAL transition which has formed the foundation of our understanding of quantum interference corrections in systems with strong symmetry breaking SOC \cite{Hikami1980,Caviglia2010,Araki2014,Kawaguchi1978}. We therefore start this section by building a physically intuitive picture as to why this reversal of transition order occurs. Afterwards, we will discuss the optimal experimental conditions to observe the novel effect.
	
	We begin by briefly recalling how the WL-to-WAL transition happens in continuous disordered metals, wherein only the spin DOF is present. Following Bergmann's interpretation \cite{Bergmann1982b,Bergmann1983}, the initial spin state of an electron, $\ket{S}$, is scattered (rotated) into some new state, $\ket{S'}$, by traversing the self-intersecting loop in one direction. This scattering may be written in terms of some general rotation matrix, $R_{s}$, such that $\ket{S'} = R_{s}\ket{S}$. By travelling along the self-intersecting loop in the opposite direction, the electron is scattered into the state $\ket{S''} = R_{s}^{-1}\ket{S}$. The overlap of these final states is given by $\langle S'' | S' \rangle = \bra{S} R_{s}^{2} \ket{S}$. This rotation of the spin is a ubiquitous manifestation of Rashba-type SOC, wherein spin and momentum are inherently linked, thus generating spin relaxation through random scattering events, irrespective of the impurity species.
	
	Now, for sufficiently strong SOC, the final spin states form a statistical ensemble and hence we must take an average value of the state overlap over all possible final spin states. Using the general rotation matrix \cite{FeynmanV3}
	\begin{equation}
		R_{s} = \begin{pmatrix}
			e^{i(\theta+\psi)/2} \cos\left(\frac{\phi}{2}\right) & i e^{-i(\theta-\psi)/2} \sin\left(\frac{\phi}{2}\right) \\
			i e^{i(\theta-\psi)/2} \sin\left(\frac{\phi}{2}\right) & e^{-i(\theta+\psi)/2} \cos\left(\frac{\phi}{2}\right)
		\end{pmatrix},
	\end{equation}
	where $\phi$, $\theta$, and $\psi$ are the Euler angles, and $\ket{S} = (a,b)^{\text{T}}$, we find that $\overline{\bra{S} R_{s}^{2} \ket{S}} = -1$, where the overline denotes averaging over the Euler angles. This average assumes that the electron's spin is able to explore the entire Bloch sphere, and hence the averages implement a uniform distribution. Therefore, the self-intersecting path reduces the probability for a particle to be found at the intersection point and so leads to anti-localisation.
	
	In the case of graphene without Rashba SOC, we replace the spin states with pseudospin states ($\ket{\Sigma}$, $\ket{\Sigma'}$, and $\ket{\Sigma''}$), whilst the rotation matrix, $R_{\sigma}$, instead acts on the pseudospin DOFs. The final pseudospin states now form a statistical ensemble, so we again take the average of their overlap. Clearly $\overline{\langle \Sigma'' | \Sigma' \rangle} = -1$.
	
	Finally, when Rashba SOC is introduced into the graphene system, the initial and final states are now products of pseudospin and spin. The initial state is given by $\ket{\Sigma,S} = \ket{\Sigma} \otimes \ket{S}$, whilst the final states are given by $\ket{\Sigma',S'} = \ket{\Sigma'} \otimes \ket{S'} = R_{\sigma}\ket{\Sigma} \otimes R_{s}\ket{S}$ and $\ket{\Sigma'',S''} = \ket{\Sigma''} \otimes \ket{S''} = R_{\sigma}^{-1}\ket{\Sigma} \otimes R_{s}^{-1}\ket{S}$. The overlap of these states is then simply $\langle \Sigma'',S'' | \Sigma',S' \rangle = \bra{\Sigma} R_{\sigma}^{2} \ket{\Sigma} \bra{S} R_{s}^{2} \ket{S}$. The pseudospin overlap may be averaged over the pseudospin Euler angles regardless of the spin. In the case of strong SOC, we also average the spin overlap over the spin Euler angles, and so the statistically averaged total overlap is  $\overline{\langle \Sigma'',S'' | \Sigma',S' \rangle} = \overline{\bra{\Sigma} R_{\sigma}^{2} \ket{\Sigma}} \, \overline{\bra{S} R_{s}^{2} \ket{S}} = -1 \times -1 = 1$. This simple yet general argument therefore predicts the quantum corrections present in ultra-clean graphene with strong Rashba coupling to be localisation in nature, reconciling perfectly with the non-perturbative calculations of the previous sections.
	
	From Fig. \ref{Quantum_corrections_plots}, we can see that for sufficiently strong intervalley scattering, the transition is reversed back to a WL-to-WAL transition. This can also be understood using the above picture of state overlap. With the inclusion of strong intervalley scattering, the electron eigenstates acquire another state label in the form of isospin, $\tau$. Hence, the averaged overlap becomes $\overline{\langle \tau'',\Sigma'',S'' | \tau',\Sigma',S' \rangle} = (-1)^{3} = -1$ for graphene systems with strong Rashba coupling and strong intervalley scattering.
	
	In addition to these SOC-driven transitions, we can encourage further change in the system's nature of localisation by varying the carrier concentration of graphene. We find that by decreasing the Fermi energy, the system moves towards the opposite phase, regardless of how its initial (anti-)localising state was established. Lowering the Fermi energy further will push the system closer to the charge neutrality point, which results in diagrams with other forms of impurity crossings (i.e. not maximally crossed) that will begin to play an important role and hence is beyond the formalism of our theory. For larger SOC strengths the $\varepsilon$-driven transition occurred at larger Fermi energies, irrespective of the intervalley scattering rate, compared to lower SOC strengths. The only notable change induced by a larger value of $\tau_{iv}^{-1}$ was a reduction in the Fermi energy at which the transition occurred for a given value of $\alpha$. We provide plots of the $\varepsilon$ dependence of $\Delta\sigma$ in \href{http://dx.doi.org/10.1038/s42005-022-01066-z}{Supplementary Note 4} for different combinations of $\alpha$ and $\tau_{v}/\tau_{iv}$.
	
	Let us now discuss the manifestation of entanglement in the Rashba-coupled graphene system. We remind the reader that such systems have been realised in graphene-TMD bilayers \cite{Wang2016,Yang2016,Yang2017,Volkl2017,Wakamura2018}, in which low-temperature magnetotransport measurements inferred the presence of an SOC ($\sim 0.1-1$meV) comparable to the quasiparticle broadening, $\eta$, hence placing the systems in the moderate to strong SOC region. In these experiments, only WAL has been observed and not WL. This is to be expected in regular disordered graphene systems due to the significant intervalley scattering rate, which fits in with the above spin scrambling picture. We therefore focus on the ultra-clean case where intervalley scattering can be neglected in the following discussion.
	
	In the absence of SOC, the transport modes are well defined in terms of effective pseudospin singlets and triplets, demonstrating an entanglement between the two particles of the Cooperon. However, when the SOC is turned on, the transport modes become combinations of pseudospin and spin singlets and triplets. These modes cannot be written purely as product states of spin and pseudospin, even for weak SOC, though they may be approximated as such when in the weak SOC limit ($\alpha \ll \eta$). Moving towards the strong SOC limit ($\alpha \gtrsim 1.1\eta$), we find that the transport modes cannot be treated as approximate product states and hence pseudospin-spin entanglement is ultimately inescapable. Different from bare graphene without SOC and 2DEGs with SOC, where entanglement manifests solely between different particles in the triplet channels, here the entanglement is also between different DOFs (i.e. mode entanglement). This mode entanglement leads to the anti-localisation states becoming less important as the Rashba coupling strength becomes well resolved within the quasiparticle broadening (i.e. when $\alpha \sim \eta$). Consequently, the reduction in significance of these states allows the localising $ss$ state to dominate and dictate the nature of the quantum correction.
	
	To-date, this WAL-to-WL transition due to increasing SOC strength has not been seen, but rather an impurity driven WL phase in the absence of SOC has been observed instead \cite{Morozov2006,Tikhonenko2008,Ki2008,Pezzini2012}. This can be attributed to the need for an ultra-clean system with minimal intervalley scattering; current experiments give $\tau_{iv}^{-1} \sim 0.7\,$meV and $\tau_{iv}^{-1} > \tau_{v}^{-1}/10$ for typical graphene sheets \cite{Pezzini2012,Yan2016}. We therefore propose that an experiment performed on ultra-clean graphene-based vdW heterostructures will observe an SOC driven WAL-to-WL transition, provided the Rashba SOC is well established in the band structure.
	
	The experimental conditions required to observe the new pseudospin-spin entanglement-driven WL phase predicted in our work could be inferred from studies similar to refs. \cite{Tikhonenko2008,Tikhonenko2009}. In the first of these experiments Tikhonenko et. al. showed that $\tau_{iv}^{-1}$ could be controlled by changing the Fermi energy via a back-gate voltage (see the Supplementary Material of Tikhonenko et. al. \cite{Tikhonenko2008}). Specifically, they showed that $\tau_{iv}^{-1}$ increased with an increase in Fermi energy while suitably far from the Dirac point. Likewise, $\tau_{v}^{-1}$ also exhibited this behaviour, though the exact functional dependence upon the Fermi energy appeared to differ between the two scattering rates. Following on from this, Tikhonenko et. al. \cite{Tikhonenko2009} demonstrated that, at a fixed temperature, lowering the Fermi energy allowed one to decrease $\tau_{iv}^{-1}$ such that a WAL phase was exhibited by their graphene samples. We therefore posit that the ratio $\tau_{v}/\tau_{iv}$ is small enough to consider the system as being ultra-clean in this scenario. Thus, we expect that a similar procedure may be implemented in back-gated graphene-TMD bilayers to enable the observation of the non-perturbative WAL-to-WL transition predicted in this work.
	
	One of the most striking features, in addition to the WAL-to-WL transition, is the appearance of turning points in the quantum corrections to the electrical conductivity, see Fig. \ref{Quantum_corrections_plots}. For the case of no intervalley scattering rate, we see an initial increase in the quantum corrections, before reaching a clear maximum in $\Delta\sigma$, as the SOC strength is increased. Further increasing $\alpha$, we see that the quantum corrections decrease rapidly towards negative values. In contrast, for significant intervalley scattering rates, the reverse is seen. This is a clear indication of the strong pseudospin-spin entanglement we have included in our analysis. We therefore emphasise that a complete understanding of SOC effects in high-mobility graphene-based vdW heterostructures requires the use of non-perturbative methods in place of the pertubative approaches commonly used to determine the Cooperon \cite{Hikami1980,Suzuura2002,McCann2006,Ochoa2014,Ilic2019,Araki2014}.
	
	Finally, let us briefly discuss how to observe this change in localisation behaviour in graphene-based vdW heterostructures in experiment. Bilayer graphene on TMD systems (BG-TMD) are ideal for probing this transition, where a perpendicular electric field can be applied to tune the effective Rashba-coupling experienced by the graphene layer adjacent to the TMD \cite{Gmitra2017,Khoo2017}. What makes BG-TMD so ideal is the range over which the SOC can be tuned, allowing for the exploration of a large region of the curves shown in Fig. \ref{Quantum_corrections_plots}. We do note that BG-TMD samples will most likely exhibit a WL-to-WAL transition, due to possessing a $2\pi$ Berry phase \cite{Kechedzhi2007}. However, we expect the same non-perturbative features to still be present in the dependence of $\Delta\sigma$ upon $\alpha$ as those we've predicted in ultra-clean graphene monolayers.
	
	\section*{Data Availability}
	The numerical data used in the generation of Fig. \ref{Quantum_corrections_plots} can be found at: Frederico Sousa, David T. S. Perkins, and Aires Ferreira. (2022). Weak Localisation Driven by Pseudospin-Spin Entanglement: Code and Data. Zenodo. \href{https://doi.org/10.5281/zenodo.7152353}{https://doi.org/10.5281/zenodo.7152353}
	
	\section*{Code Availability}
	The code used for this project is available at: Frederico Sousa, David T. S. Perkins, and Aires Ferreira. (2022). Weak Localisation Driven by Pseudospin-Spin Entanglement: Code and Data. Zenodo. \href{https://doi.org/10.5281/zenodo.7152353}{https://doi.org/10.5281/zenodo.7152353}
	
	\bibliographystyle{unsrt}

	
	\section*{Acknowledgements}
	We would like to thank T. Wakamura for their comments on this manuscript. \textbf{Funding:} D.T.S.P. and A.F. acknowledge funding from the Royal Society (Grant No. URF\textbackslash R\textbackslash191021 and RF\textbackslash ERE\textbackslash 210281).
	
	\section*{Author Contributions}
	A.F. and F.S. derived the generalised non-perturbative expression for the Cooperon self-energy. F.S. performed the field-theoretic calculations detailed in this manuscript. D.T.S.P. provided the spin scrambling interpretation of the WAL-to-WL transition detailed here, and wrote the manuscript with input from all authors. A.F. supervised this project and manuscript writing, providing help and advice on how to move forward.
	
	\section*{Competing Interests}
	The authors declare that they have no competing interests.

\end{document}